\documentclass[12pt,a4paper]{article}
\usepackage[T1]{fontenc}
\usepackage{geometry}
\usepackage{graphicx}
\usepackage{multicol}
\usepackage{comment}

\usepackage{amssymb,amsopn,amsmath,amstext,amsfonts,algorithmic,psfrag}
\usepackage{amssymb}
\usepackage{enumitem}
\usepackage{mathtools}
\usepackage{hyperref}
\usepackage[round]{natbib}
\usepackage{authblk}

\usepackage{eso-pic}
\newcommand\AtPageUpperMyright[1]{\AtPageUpperLeft{%
 \put(\LenToUnit{0.5\paperwidth},\LenToUnit{-1cm}){%
     \parbox{0.5\textwidth}{\raggedleft\fontsize{9}{11}\selectfont #1}}%
 }}%
\newcommand{\conf}[1]{%
\AddToShipoutPictureBG*{%
\AtPageUpperMyright{#1}
}
}

\title{\LARGE \bf
Game-Theoretic Model Based Resource Allocation During Floods
}

\author[1]{Rudrashis Majumder}
\author[2]{Rakesh R Warier}
\author[3]{Debasish Ghose}

\affil[1]{Ph.D. Student, Department of Aerospace Engineering, Indian Institute of Science, Bangalore, 
		{\tt\small rudrashismajumder@gmail.com}}

\affil[2]{Post-doctoral Fellow, Department of Aerospace Engineering, Indian Institute of Science, Bangalore,
		{\tt\small rakeshwarier@iisc.ac.in}}

\affil[3]{Professor, Department of Aerospace Engineering, Indian Institute of Science, Bangalore, 
		{\tt\small dghose@iisc.ac.in}}		
		
\date{}                     
\begin{document}

\maketitle

\conf{$4^{th}$ World Congress on  Disaster Management\\ IIT Bombay, Mumbai, India, 29 Jan-01 Feb 2019}

\begin{abstract}

For multiple emergencies caused by natural disasters, it is crucial to allocate resources equitably to each emergency location, especially when the availability of resources is limited in quantity. This paper has developed a multi-event crisis management system using a non-cooperative, complete information, strategic form game model. In the proposed system, each emergency event is assumed to occur in different locations simultaneously. These locations are represented as the players in the game, competing with the other players for an optimal allocation of scarce resources available at different resource stations. The players incur a non-monetary cost for obtaining resource units. The objective of the proposed game is to derive optimal strategies for an effective and fair allocation of resources to the respective players.

\end{abstract}


\section{Introduction}\label{sec:intro}

The efficient and quick response towards natural disasters involves a set of activities that help in reducing the after-effect of the disasters. The occurrence of floods causes casualties and infrastructure damage. Search and rescue operations, evacuation of affected people, medical assistance, providing food, pure water, and other essential materials are a few activities that should start as soon as possible to minimize the damage caused by a flood. Delayed and unplanned response by the disaster management authority makes the situation worse for the inhabitants of the flood-affected area. 
 
Various types of resources are required in the emergency sites to deal with the circumstances caused by a flood. Faster rescue operations, food distribution, medicine, etc., are critical requirements in flood-affected areas. When more than one crisis occurs at different locations simultaneously due to a natural disaster, each crisis location needs a specific type and certain amount of resources in a time-overlapping manner. So, it is vital to fairly allocate resource units to each crisis location, particularly when the available resources are limited in quantity. 

Game theory is concerned with the analysis of strategic relations between rational agents and is functional in a variety of applications \citep{roger1991game}. Pioneering findings of \cite{von2007theory, nash1951non} have made game theory an irreplaceable tool for decision-making. Multiple collaborators, including various government and non-government agencies, private organizations, citizens, and others, must interact and coordinate for achieving effective disaster response. Since disaster management consists of strategic interactions of different decision-makers, game theory becomes a viable option for modeling disaster response. Recent research in disaster management indicates the growing use of game-theoretic models \citep{seaberg2017review}. \cite{gupta2006social, ranganathan2007automated, gupta2007multievent, wang2009game} use a non-cooperative game model where the game is played between crisis locations as each location aspires to have the most resources possible. The authors consider a multi-event emergency management system based on a strategic game where the cost matrix for the game is computed based on the criticality of the events and the response time of the resources to reach the crisis location. \cite{ranganathan2007automated} refer to social fairness as the metric for the utility model. \cite{wang2009game} use a complete information game model, and the Nash equilibrium is attained by using an improved Ant Colony optimization-based computational algorithm. A sequential game proposed by \cite{yang2012emergency} considers decision-makers and emergency as two players. A dynamic game model with incomplete and complete information is used in this paper. There is a substantial amount of research work that addresses various problems related to flood-affected areas. \cite{Balasubramaniam2019LiDAR} addresses the objective of land-water boundary identification based on information from LiDAR mounted on UAV.  \cite{shriwastav2019images} present an approach for detection of land-water boundaries using UAV-acquired images of the area. \cite{Kashyap2019urban} present an approach to reroute the resource traffic based on instantaneous inundation information of the flooded region, and \cite{ravichandran2019searching} present an approach for detection and tracking of survivors in the flooded region by using UAV-based information. 

The difficulty of responding to several crises that occur at the same time by allocating appropriate resources is examined in this study. A real-life scenario has been converted into a non-cooperative game to find the optimal solution to a resource allocation problem in a situation where resource quantity is limited. The floods in the Indian state of Kerala in 2018, which stranded thousands of people in several areas, inspired our research. It was critical to allocate resources such as rescue helicopters, boats, relief supplies, and medical assistance to different crisis regions. We analyze the distribution of rescue helicopters as a non-cooperative game between crisis areas and propose an equilibrium solution as an efficient resource distribution method.  

The paper models the problem of resource allocation with insufficient resource units using a non-cooperative game. A novel aspect of the proposed approach includes a strategy to select an equilibrium solution when there are multiple Pure Strategy Nash equilibria (PSNE) of the game. Additionally, our choice of the cost function can allocate the responsibility of conflict caused by individual players by the demands they make on available resources. 

The paper is organized as follows. Section \ref{sec:game} introduces the basic concepts of game theory. The concept of equilibrium solution and two important criteria useful for the selection of a particular equilibrium solution from a set of equilibria are defined. The resource allocation game from a real-life scenario is modeled here. The choice and decisions to be taken by the players are explained in Section \ref{sec:choice}. In Section \ref{sec:disc}, we discuss the work and the findings briefly. Section \ref{sec:concl} gives the concluding remarks.

\section{Game Formulation and Nash equilibrium}\label{sec:game}

Game theory studies the strategic interaction between several rational and intelligent decision-makers. In the case of non-cooperative games, players will try to improve their utility or payoff (or reduce the cost incurred by them) individually. Here, we use a non-cooperative strategic form game model. These are also called matrix games, and we represent these types of games through a payoff or cost matrix. 

A strategic form game can be represented as $ G = \left\langle N, {(S_i)}_{\forall i \in N}, {(U_i)}_{\forall i \in N} \right\rangle $ where
\begin{enumerate}[label=(\roman*)]
	\item  $ N = \{ 1, 2, ..., n\} $ is a set of players
	\item $ S_1, S_2, ..., S_n $ are the strategy sets of the players $i = \{1, 2, ..., n\}$ respectively
	\item $ U_i: S_1 \times S_2 \times ... \times S_n \rightarrow \mathbb{R} $ for $ i = {1, 2, ..., n}$ are  the utility, or cost functions.
\end{enumerate}

Given a strategic form game $ G = \left\langle N, (S_i), (U_i) \right\rangle $, a strategy profile $ s^* = ({s_1}^*, {s_2}^*, ..., {s_n}^*) $ is called a pure strategy Nash equilibrium (PSNE) of game $ G $ if
\begin{equation}\label{eq:1}
U_i({s_i}^*, s_{-i}) \leqslant U_i(s_i, s_{-i}), \forall s_i \in S_i, \forall i \in N
\end{equation}
Actions of the players other than the $ i $-th player are denoted as $  s_{-i} $.

\subsection{Payoff and Risk Dominance}\label{pdrd}

A pure-strategy Nash equilibrium (PSNE) is an action profile such that no single player can obtain a better incentive by deviating unilaterally from this profile. If solving a game leads to multiple PSNEs, we have to choose one PSNE out of the set for implementation. To find a unique PSNE, we leverage two properties: payoff and risk dominance. The matrix below shows the cost matrices for a two player-two strategy game. The row player P1 chooses between rows $ X $ and $ Y $. Player P2 is the column player choosing columns $ X $ and $ Y $.
\begin{table}[h!]
	\begin{center}
		\begin{tabular}{ |p{0.5cm}|p{1.0cm}|p{1.0cm}| }
			\hline
			& $ X $ &  $ Y $ \\
			\hline
			$ X $ &  $ A, a $ & $  C, b $ \\
			\hline
			$ Y $ &  $ B, c $ & $  D, d $ \\
			\hline
		\end{tabular}
	\end{center}
\end{table}

Assuming that the above game has two PSNE: $ (X, X) $ and $ (Y, Y) $, their corresponding outcomes are $ (A, a) $ and $ (D, d) $. With $ A \leqslant D $ and $ a \leqslant d $, and at least one strict equality, the PSNE $ (X, X) $ is payoff dominant \citep{harsanyi1988general}. 

Now, if $ (A - B)(a - b) \geqslant (D - C)(d - c) $ then $ (X, X) $ is the risk dominant PSNE with the outcome $ (A, a) $ \citep{harsanyi1988general}. Otherwise $ (Y, Y) $ is the risk dominant solution.  $ (A - B)(a - b) $ and $ (D - C)(d - c) $ are called the Nash product, or the product of gains for unilateral deviations of the equilibrium $ (X, X) $ and $ (Y, Y) $, respectively.

\subsection{A Realistic Example}
There was a devastating flood that occurred in Kerala during July-August 2018 \citep{KeralaFloodsMishra2018}. The main reason behind this was unusually high rainfall during the monsoon season. The rivers overflowed, and water was released from the dams. Some districts in the hilly region were also paralyzed by the repeated occurrence of landslides caused by floodwaters eroding the landmass. Most parts of Kerala were under high alert for an extended period of time. The rescue work was conducted by the Military, Army, Navy, NDRF, NDMA, Coastal Guards, BSF, and many non-government volunteers. Nearly one million people were evacuated from flood-affected areas to the relief camps. Army and Navy used their helicopters for the rescue and relief operation in flood and landslide affected areas. However, multiple disaster events co-occurring at different locations made the resource allocation and other disaster responses extremely difficult.

\begin{figure*}
    \centering
    \includegraphics[width = 0.8\textwidth]{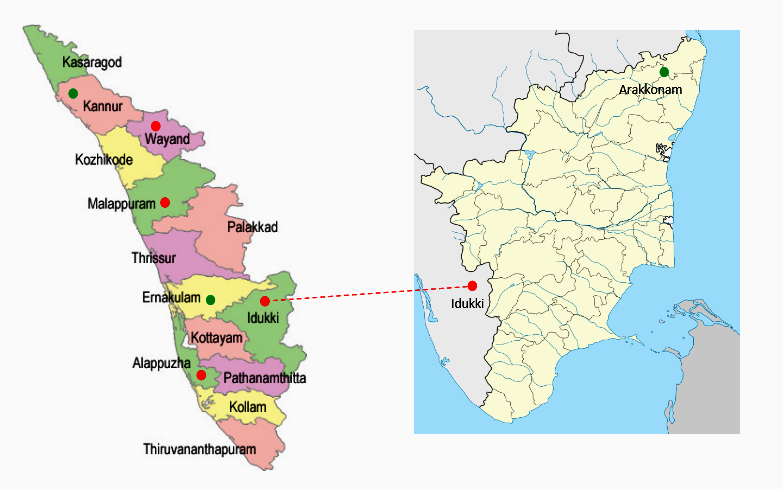}
    \caption{Crisis locations (red dots) and Resource stations (green dots) (Image from Wikimedia commons)}
    \label{fig:my_label}
\end{figure*}

As a case study of our framework, we consider a few districts of Kerala as the players affected by the flood and landslides (for example, Alappuzha, Idukki, Malappuram, and Wayanad). Here onwards, we will use the words ``crisis locations'' and ``players'' synonymously. Alappuzha and Malappuram were affected by the flood, while Idukki and Wayanad were affected mainly by landslides. Primarily, Navy helicopters were used for rescue and relief operations. Helicopters getting dispatched from different Naval basecamps for rescue operations are considered as the resources for which the players or disaster locations compete. We use the words ``resource'' and ``helicopters'' synonymously in our discussion. \par
We consider two Naval Base stations Cochin (RS1) and Ezhimala (RS2), as the resource stations in Kerala. However, as shown in Table~\ref{table:a}, Arakkonam (RS3) in Tamilnadu is the third resource station that will act as a backup and send helicopters in case the resources are available in Kerala resource stations RS1 and RS2 are not sufficient to meet the need of the players. We formulate this problem as a noncooperative game and show that the solution to the game leads to a fair allocation of resources.  

\begin{table}[h!]
	\begin{center}
		\caption{Resource stations (RS)}
		\begin{tabular}{|c|c|} 
			\hline
			\bf Naval Base Stations in Kerala & \bf Naval Base Stations in Tamilnadu \\ 
			\hline\hline
			Cochin (RS1) in Ernakulam & {Arakkonam (RS3)}  \\ 
			Ezhimala (RS2) in Kannur & \\
			\hline
		\end{tabular}\label{table:a}	
	\end{center}
\end{table}


The requirement of the helicopters for a particular crisis location is calculated as a function of:
\begin{enumerate}
	\item Area of the crisis location, ( $A$ ).
	\item Criticality of the crisis location ($ \rho $).
	\item Resources needed per unit area (sq. km) of the crisis location ($N$).
\end{enumerate}
The number of helicopters needed for a particular crisis location is calculated as the product $ A \rho N $. In the crisis locations, we assume that the number of helicopters required is equal to one per $400$ sq. km. The total number of helicopters required for the players is calculated in Table \ref{table:b}.
\begin{table}[h!]  
	\begin{center}
		\caption{Computation of requirements of helicopters}
		\begin{tabular}{ |p{3.5cm}|p{2.0cm}|p{2.5cm}|p{2.2cm}|  }
			\hline
			\bf Crisis location/ Players & \bf Area (in sq. km.) & \bf Criticality factor & \bf Helicopters needed \\
			\hline\hline
			P1: Alappuzha & $1414$ & $0.8$ & $ 3 $ \\
			\hline
			P2: Idukki & $4358$ & $0.6$ & $ 7 $ \\
			\hline
			P3: Malappuram & $3550$ & $0.75$ & $ 7 $ \\
			\hline
			P4: Wayanad & $2132$ & $0.65$ & $ 4 $ \\
			\hline
		\end{tabular}\label{table:b}
	\end{center}
\end{table}

The players Alappuzha and Idukki are closer to the resource station RS1 whereas players Malappuram and Wayanad are closer to the resource station RS2. Table \ref{table:c} shows the sorted order of distances of resource stations from the players.

\begin{table}[h!]
	\begin{center}
		\caption{Distances from the resource stations and Sorted order of distances}
		\begin{tabular}{ |p{1.5cm}|p{2cm}|p{2cm}| p{2cm}| p{4cm}| }
			\hline
			\bf Players & \bf Approx. distance from RS1 (in km) & \bf Approx. distance from RS2 (in km) & \bf Approx. distance from RS3 (in km) & \bf Sorted order of resource stations in terms of distance \\
			\hline\hline
			P1 & $70$ & $375$ & $650$ & RS1 $<$ RS2 $<$ RS3 \\
			P2 & $100$ & $350$ & $500$ & RS1 $<$ RS2 $<$ RS3 \\
			P3 & $186$ & $150$ & $560$ & RS2 $<$ RS1 $<$ RS3 \\
			P4 & $200$ & $130$ & $530$ & RS2 $<$ RS1 $<$ RS3 \\
			\hline
		\end{tabular}\label{table:c}
	\end{center}
\end{table}

However, the requirement of resources may not be satisfied from the nearest resource stations. We assume that the number of helicopters available to the resource stations is as given in Table \ref{table:d}.
\begin{table}[h!]
	\begin{center}
		\caption{Availability of helicopters in the resource stations}
		\begin{tabular}{ |p{4cm}|p{2.2cm}| }
			\hline
			\bf Naval Air Stations & \bf Units available \\
			\hline\hline
			Cochin (RS1) & $8$ \\
			Ezhimala (RS2) & $9$ \\
			\hline
			Arakkonam (RS3) & $6$ \\
			\hline
		\end{tabular}\label{table:d}
	\end{center}
\end{table}

From the availability of resources in the considered resource stations, we can see a conflict situation in sharing the resources. From Table \ref{table:e}, we see that the total requirement of the players exceeds the number of available resources in RS1 and RS2. Since P1 and P2 are nearer to RS1, first, they will try to get their needed resources from RS1. If the resources available to RS1 are not sufficient for them, they will see the status in RS2. Since RS2 is already in a resource-constraint situation (to serve P3 and P4), they will take RS3 as a backup resource station.  Similarly, P3 and P4 will first try to get their resources from RS2 and then from RS3. As Table \ref{table:e} shows, RS3 has adequate resources to fulfill the deficiencies of the players.

\begin{table}[h!]
	\begin{center}
		\caption{Conflict between total requirement and availability}
		\begin{tabular}{ |p{4cm}|p{1.0cm}|p{1.0cm}|p{1.0cm}| }
			\hline
			\bf Players& \bf RS1 & \bf RS2 & \bf RS3 \\
			\hline\hline
			\bf P1 & $3$ & $0$ & $0$ \\
			\hline
			\bf P2 & $7$ & $0$ & $0$ \\
			\hline
			\bf P3 & $0$ & $7$ & $0$ \\
			\hline
			\bf P4 & $0$ & $4$ & $0$ \\
			\hline\hline
			\bf Total requirement & $10$ & $11$ & $0$ \\
			\hline
			\bf Availability & $8$ & $9$ & $6$ \\
			\hline
		\end{tabular}\label{table:e}
	\end{center}
\end{table}

For effective disaster response, the players need a certain number of helicopters. However, helicopters in the nearest resource stations are insufficient to fulfill the requirement of both the players. In this situation, P1 and P2 play a two-player game to divide resources available in RS1 fairly. In the same way, P3 and P4 play a two-player game to divide resources of RS2. A general cost function model is proposed for the players, which calculates the cost or penalty charged for requesting several resource units. The entries of the game matrices are derived from this cost function. Each player has the intention to reduce the cost being charged on them for their own choices. The Nash equilibrium of the game identifies an optimal strategy for the players to choose.

\subsubsection{\bf Cost in terms of response time}

Mostly Sea King helicopters from the Indian Navy were used for the rescue operations, with an average speed of 210 km per hour. We can calculate the time required for the helicopters to reach the crisis location from the resource stations from this data. Table \ref{table:f} gives the cost in terms of response time. If the resource is coming from a nearer resource station, the response time will be less. If the player is forced to bring resources from a distant location, the response time will increase accordingly.

\begin{table}[h!]
	\begin{center}
		\caption{Cost in terms of response time}
		\begin{tabular}{ |p{1.5cm}|p{1.8cm}|p{2cm}|p{2.0cm}| }
			\hline
			\bf Players & \bf Resource Stations & \bf Distance (km) & $ \bf P_t (hr.) $ \\
			\hline\hline
			P1 & RS1 & $70$ &  $ 0.33 $\\
			\hline
			P1 & RS3 & $650$ & $ 3.1 $ \\
			\hline
			P2 & RS1 & $100$ & $  0.48 $ \\
			\hline
			P2 & RS3 & $500$ & $ 2.4 $ \\
			\hline\hline
			P3 & RS2 & $150$ & $ 0.71 $\\
			\hline
			P3 & RS3 & $560$ & $ 2.67 $ \\
			\hline
			P4 & RS2 & $135$ & $ 0.64 $ \\
			\hline
			P4 & RS3 & $530$ & $ 2.52 $ \\
			\hline
		\end{tabular}\label{table:f}
	\end{center}
\end{table}

\subsubsection{\bf Cost in terms of fuel consumption}

For Sea King Helicopters, 3700 liters of fuel gives a range of 1500 km. Thus fuel consumption rate is approximately 0.0025 kilolitre/km. Table \ref{table:g} shows the fuel consumption amount for bringing just one helicopter from the respective resource locations. 

\begin{table}[h!]
	\begin{center}
		\caption{Cost in terms of fuel consumption}
		\begin{tabular}{ |p{1.5cm}|p{1.8cm}|p{2cm}|p{2.2cm}|  }
			\hline
			\bf Player & \bf Resource stations &  \bf Distance (in km.) &  $ \bf P_c (kilolitre) $ \\
			\hline\hline
			P1 & RS1 & $70$ & $ 0.175 $ \\
			\hline
			P1 & RS3 & $650$ & $ 1.625 $ \\
			\hline
			P2 & RS1 & $100$ & $ 0.250 $ \\
			\hline
			P2 & RS3 & $500$ & $ 1.250 $ \\
			\hline\hline
			P3 & RS2 & $150$ & $ 0.375 $ \\
			\hline
			P3 & RS3 & $560$ & $ 1.4 $ \\
			\hline
			P4 & RS2 & $135$ & $ 0.3375 $ \\
			\hline
			P4 & RS3 & $530$ & $ 1.325 $ \\
			\hline
		\end{tabular}\label{table:g}
	\end{center}
\end{table}

\subsubsection{\bf Penalty based on their mutual actions}

Players will always try to take the maximum possible resources from the nearest resource stations. However, since this may demand more than the number of available resources, another penalty is charged on the players so that they are not inclined to take all their required resources from the nearest resource station. For a two-player game, (\ref{eq:2}) and (\ref{eq:3}) gives this penalty for P1 and P2. 
\begin{align}
P_{l_1} =&~f(A_1,d_1,d_2) \bigg[ \frac{n_1-d_1}{n_1} \bigg] + \frac{n_1-d_1}{n_1}  + \frac{n_2-(A_1-d_1)}{n_2} + g(A_1,d_1,d_2)\bigg[ \frac{\frac{d_1}{n_1}}{\frac{d_1}{n_1}+\frac{d_2}{n_2}} \bigg] \label{eq:2} \\
P_{l_2} =&~f(A_1,d_1,d_2) \bigg[\frac{n_2-d_2}{n_2} \bigg]  + \frac{n_2-d_2}{n_2}  + \frac{n_1-(A_1-d_2)}{n_1} + g(A_1,d_1,d_2) \bigg[\frac{\frac{d_2}{n_2}}{\frac{d_1}{n_1}+\frac{d_2}{n_2}} \bigg] \label{eq:3}
\end{align}

where $f(.)$ and $g(.)$ functions are defined as follows:

\begin{eqnarray}
f(x,y,z) & = & \begin{cases}
x-y-z & \text{\text{if }}(x-y-z)\geq0\\
0 & \text{otherwise}
\end{cases} \label{eq:fdef}\\ 
g(x,y,z) & = & \begin{cases}
y+z-x & \text{\text{if }}(x-y-z)\leq0\\
0 & \text{otherwise}
\end{cases} \label{eq:gdef}
\end{eqnarray}

Here, $ n_1 $ and $ n_2 $ are number of resources needed by players P1 and P2, respectively. $ d_1 $ and $ d_2 $ are number of resources requested by players P1 and P2, respectively. The number of resources available to RS1 is denoted by $ A_1 $. \par

In (\ref{eq:2}) and (\ref{eq:3}), the first term corresponds to the under-utilization by the players, penalizing players when the total demand is less than the available number of resources. This term is nonzero when the total number of resource units requested by two players is less than the availability. The second term corresponds to the dissatisfaction due to the player's own choice. The second term penalizes the player when the demand $d_i$ is less than $n_i$. The third term indicates how much restriction one player puts on the other player. The fourth term penalizes the conflict created by two players. This term becomes nonzero if the total number of units requested by the players exceeds the availability. The fourth term tries to allocate the responsibility of conflict between players fairly.

For P3 and P4 in RS2, we get:

\begin{align}
P_{l_3} =&~f(A_2,d_3,d_4) \bigg[\frac{n_3-d_3}{n_3} \bigg] + \frac{n_3-d_3}{n_3}  + \frac{n_4-(A_2-d_3)}{n_4} + g(A_2,d_3,d_4)\bigg[ \frac{\frac{d_3}{n_3}}{\frac{d_3}{n_3}+\frac{d_4}{n_4}} \bigg] \label{eq:5} \\
P_{l_4} =&~f(A_2,d_3,d_4) \bigg[ \frac{n_4-d_4}{n_4} \bigg]  + \frac{n_4-d_4}{n_4}  + \frac{n_3-(A_2-d_4)}{n_3} + g(A_2,d_3,d_4) \bigg[\frac{\frac{d_4}{n_4}}{\frac{d_3}{n_3}+\frac{d_4}{n_4}} \bigg] \label{eq:6}
\end{align}

where $f(.)$ and $g(.)$ functions are defined similarly as (\ref{eq:fdef}) and (\ref{eq:gdef}).

Here, $ n_3 $ and $ n_4 $ are the number of resources needed by players P3 and P4, respectively. $ d_3 $ and $ d_4 $ are the number of resources requested by P3 and P4, respectively. The number of resources available to RS2 is denoted by $ A_2 $. \par

\subsubsection{\bf Total cost}
The total cost incurred by the $i^\text{th}$ player is given as follows:
\begin{equation}\label{eq:7}
P_i=\alpha_t  P_{t_i} + \alpha_c  P_{c_i} + \alpha_l  P_{l_i}
\end{equation}
where, $ \alpha_t, \alpha_c, \alpha_l $ are positive scalar values and $ \alpha_t + \alpha_c + \alpha_l = 1 $. The total cost gives the entries of the payoff or cost matrices for all the players.

\section{Choice and decisions}\label{sec:choice}

From the cost matrix, we calculate the Nash equilibrium for P1 and P2 in RS1 and for P3 and P4 in RS2. All the players will have the objective to reduce the cost incurred on them. \par

Case 1: When $ \alpha_t = 0.05, \alpha_c = 0.05, \alpha_l = 0.9 $, we get multiple pure strategy Nash equilibria (Table \ref{table:h}).
\begin{table}[h!]
	\begin{center}
		\caption{Nash Equilibrium for $ \alpha_t = 0.05, \alpha_c = 0.05, \alpha_l = 0.9 $}
		\begin{tabular}{ |p{2.2cm}|p{4cm}||p{2.2cm}|p{4cm}| }
			\hline
			\bf NE at RS1 & \bf Costs on P1 and P2 & \bf NE at RS2 & \bf Costs on P3 and P4 \\
			\hline\hline
			$(1, 7)$ & $(0.92601, 0.71131)$  & $(5, 4)$ & $(0.62423, 0.3531)$  \\
			\hline
			$(2, 6)$ & $(0.68208, 0.68512)$ & $(6, 3)$ & $(0.6694, 0.59476)$  \\
			\hline
			$(3, 5)$ & $(0.30006, 0.56369)$ & $(7, 2)$ & $(0.61696, 0.74119)$  \\
			\hline
		\end{tabular}\label{table:h}
	\end{center}
\end{table}
All the solutions we obtain for this set of $ \alpha $-values are obvious solutions. Payoff and risk dominance defined in Section \ref{sec:game} helps us to choose a unique PSNE as the intended solution. In the above case, (3, 5) is both payoff and risk dominant solution for P1 and P2. However, in RS2, there is no payoff-dominant solution. However, equilibrium (5, 4) is the risk-dominant solution. \par

The interpretation of the solution is that in RS1, the optimal strategy for P1 is to choose three units of helicopters, and the optimal strategy for P2 is to choose five units of helicopters. This is the way to allocate all the eight units of helicopters available in RS1 among P1 and P2. Similarly, in RS2, with nine helicopters available, the optimal strategy for P3 is to choose five units of helicopters, and the optimal strategy for P4 is to choose four units of helicopters. RS3 being a backup resource station, will supply the additional resources needed by the players. However, since it is far away from the crisis locations, fair allocation of resources from the nearby resource location is necessary for immediate initiation of disaster response.

Case 2: When $ \alpha_t = 0.2, \alpha_c = 0.05, \alpha_l = 0.75 $, we get unique PSNE (3, 5) in RS1 and two PSNE in RS2 (Table \ref{table:i}): 
\begin{table}[h!]
	\begin{center}
		\caption{ Nash Equilibrium for $\alpha_t = 0.2, \alpha_c = 0.05, \alpha_l = 0.75 $}
		\begin{tabular}{ |p{2.2cm}|p{4cm}|p{2.2cm}|p{4cm}| }
			\hline
			\bf NE at RS1 & \bf Costs on P1 and P2 & \bf NE at RS2 & \bf Costs on P3 and P4 \\
			\hline\hline
			$(3, 5)$ & $(0.3072, 0.87798)$  & $(5, 4)$ & $(0.98137, 0.4031)$ \\
			\hline
			& & $(7, 2)$ & $(0.64911, 1.0448)$ \\
			\hline
		\end{tabular}\label{table:i}
	\end{center}
\end{table}

In RS2, none of the equilibrium solutions is payoff dominant. However, (5, 4) is the risk-dominant solution, and it risk-dominates (7, 2). We can interpret these solutions in the same way as we have done in Case 1.

Case 3: When $ \alpha_t = 0.1, \alpha_c = 0.25, \alpha_l = 0.65 $, we get unique PSNE for both RS1 and RS2 (Table \ref{table:j}): 
\begin{table}[h!]
	\begin{center}
		\caption{ Nash Equilibrium for $\alpha_t = 0.1, \alpha_c = 0.25, \alpha_l = 0.65 $ }
		\begin{tabular}{ |p{2.2cm}|p{4cm}|p{2.2cm}|p{4cm}| }
			\hline
			\bf NE at RS1 & \bf Costs on P1 and P2 & \bf NE at RS2 & \bf Costs on P3 and P4 \\
			\hline\hline
			$(3, 5)$ & $(0.3503, 1.3613)$  & $(5, 4)$ & $(1.6211, 0.57262)$  \\
			\hline
		\end{tabular}\label{table:j}
	\end{center}
\end{table}

\section{Discussions}\label{sec:disc}

The example in Section \ref{sec:game} demonstrates how a game theory-based solution can be used to solve a resource allocation problem in a resource-constrained environment after a natural disaster. The game has been formulated to study resource allocation after the Kerala flood in 2018. The crisis locations are the players, and they incur some cost for availing resources. For different values of the $ \alpha $-set in (\ref{eq:7}), we get different pure strategy Nash equilibria (PSNE). The Nash equilibrium is the best strategy for each player when the other players take their best strategies. The PSNE can be interpreted as the way of allocating resources fairly and optimally.  

However, depending on the chosen values of the $ \alpha $-set, sometimes multiple Nash equilibria may come into the picture. In the case of a single unique PSNE, the allocation of resources becomes straightforward. However, in the case of multiple PSNE, selecting a single solution is taken based on payoff dominance and risk dominance properties of the obtained equilibria. If a particular PSNE payoff-dominates the others, we select it as the desirable PSNE. If payoff dominance fails, the decision is taken based on risk dominance.

The chosen $ \alpha $-values give PSNE for all Case 1-3. However, if PSNE does not exist, we can find mixed strategy Nash equilibria (MSNE) as it will always exist \citep{nash1951non}. In the case of MSNE, the players choose their actions with some probability, not deterministically. MSNE solutions tell how the resources can be allocated to the crisis locations by imposing probabilistic notions.

\section{Conclusions}\label{sec:concl}

In this paper, a game-theoretic solution for multi-event crisis management has been proposed. This method identifies the crisis locations as the players. The resources are available in a limited quantity; hence, one player's choice directly impacts the choices made by the other players. A non-cooperative game is formulated between the concerned players for a socially fair distribution of resources. The players compete for a near-optimal allocation of resource units. The Nash equilibrium-based optimization is implemented in the game to derive a desirable allocation of resources. The entire idea has been explained using a practical example based on the recent Kerala flood in India. The results produced in this paper show that, in a resource-constrained situation, suitably designed game-theoretic algorithms can provide a potential and efficient solution for resource allocation.

\section*{Acknowledgement}
This work is supported by an EPSRC-GCRF project `Emergency Flood
		Planning and Management using Unmanned Aerial Systems' through an
		international multi-institutional grant. The authors would like to thank their
		colleagues in the EPSRC-GCRF project partner institutes, University of Exeter, Indian Institute of Science, Indraprastha Institute of Information Technology Delhi, University of Cranfield, and the Tata Consultancy Services, for their
		help and assistance.

\bibliographystyle{spbasic}
\bibliography{ref}

\end{document}